\documentstyle [twocolumn,epsf]{mn}
\newcommand{\mincir}{\raise
-2.truept\hbox{\rlap{\hbox{$\sim$}}\raise5.truept\hbox{$<$}\ }}
\newcommand{\magcir}{\raise
-2.truept\hbox{\rlap{\hbox{$\sim$}}\raise5.truept\hbox{$>$}\ }}
\newcommand{\minmag}{\raise
-2.truept\hbox{\rlap{\hbox{$<$}}\raise6.truept\hbox{$<$}\ }}
\newcommand{\be}{\begin{equation}}
\newcommand{\ee}{\end{equation}}
\newcommand{\ba}{\begin{eqnarray}}
\newcommand{\ea}{\end{esqnarray}}
\newcommand{\brr}{\begin{array}}
\newcommand{\err}{\end{array}}
\newcommand{\bc}{\begin{center}}
\newcommand{\ec}{\end{center}}

\renewcommand{\baselinestretch}{1.2}
\title[Is the CMB shift parameter connected with the growth factor?]{Is the CMB shift parameter connected with the growth of cosmological perturbations?}

\author[S. Basilakos, S. Nesseris \& L. Perivolaropoulos]
{S. Basilakos$^{1}$, S. Nesseris$^{2}$ \& L. Perivolaropoulos$^{2}$ \\
$^{1}$ Academy of Athens, Research Center for Astronomy \& Applied
  Mathematics, Soranou Efessiou 4, 11-527, Athens, Greece\\
$^2$ Department of Physics, University of Ioannina, Greece\\
}

\begin{document}

\maketitle

\begin{abstract}
We verify numerically that in the context of general relativity (GR),
flat models which have the same $\Omega_{\rm m}$ and CMB shift
parameter $R$ but different $H(a)$ and
$w(a)$ also have very similar (within less than $8\%$) growth of
perturbations even though the dark energy density evolution is quite
different. This provides a direct connection between geometrical and
dynamical tests of dark energy and may be used as a cosmological
test of general relativity.

\vspace{0.25cm}

\end{abstract}

\section{Introduction}
There is by now convincing evidence that the available high quality
cosmological data (Type Ia supernovae, CMB, etc.) are well fitted by
an emerging ``standard model''. In the context of GR this ``standard
model'', assuming flatness, is described by the Friedman equation
\be H^2(a)=\left(\frac{{\dot a}}{a}\right)^2=H_0^2\left[\Omega_{\rm
m}(a)+\Omega_{\rm DE}(a)\right] \label{fe1}\ee where $a(t)$ is the
scale factor of the universe, $\Omega_{\rm m}(a)$ is the density
parameter corresponding to the sum of baryonic and cold dark matter,
with the latter needed to explain clustering, and an extra component
$\Omega_{\rm DE}(a)$ with negative pressure called dark energy needed to
explain the observed accelerated cosmic expansion (Riess et al.
1998; Perlmutter et al. 1999; Efstathiou et al. 2002; Tegmark et al.
2004; Spergel et al. 2007;  Nesseris, \&  Perivolaropoulos 2005;
Nesseris, \&  Perivolaropoulos 2007a). During the last decade there
have been many theoretical speculations regarding the nature of the
exotic ``dark energy''. Various candidates have been proposed in the
literature, among which a dynamical scalar field acting as vacuum
energy (Ozer \& Taha 1987; Caldwell, Dave \& Steinhardt 1998;
Peebles \& Ratra 2003). Under this framework, high energy field
theories generically indicate that the equation of state of such a
dark energy is a function of the cosmic time. To identify this type
of evolution of the equation of state, a detailed form of the
observed $H(z)$ is required which may be obtained by a combination
of multiple dark energy probes. Such probes may be divided in two
classes according to the methods used to obtain $H(z)$.

\begin{figure*}
\mbox{\epsfxsize=12cm \epsffile{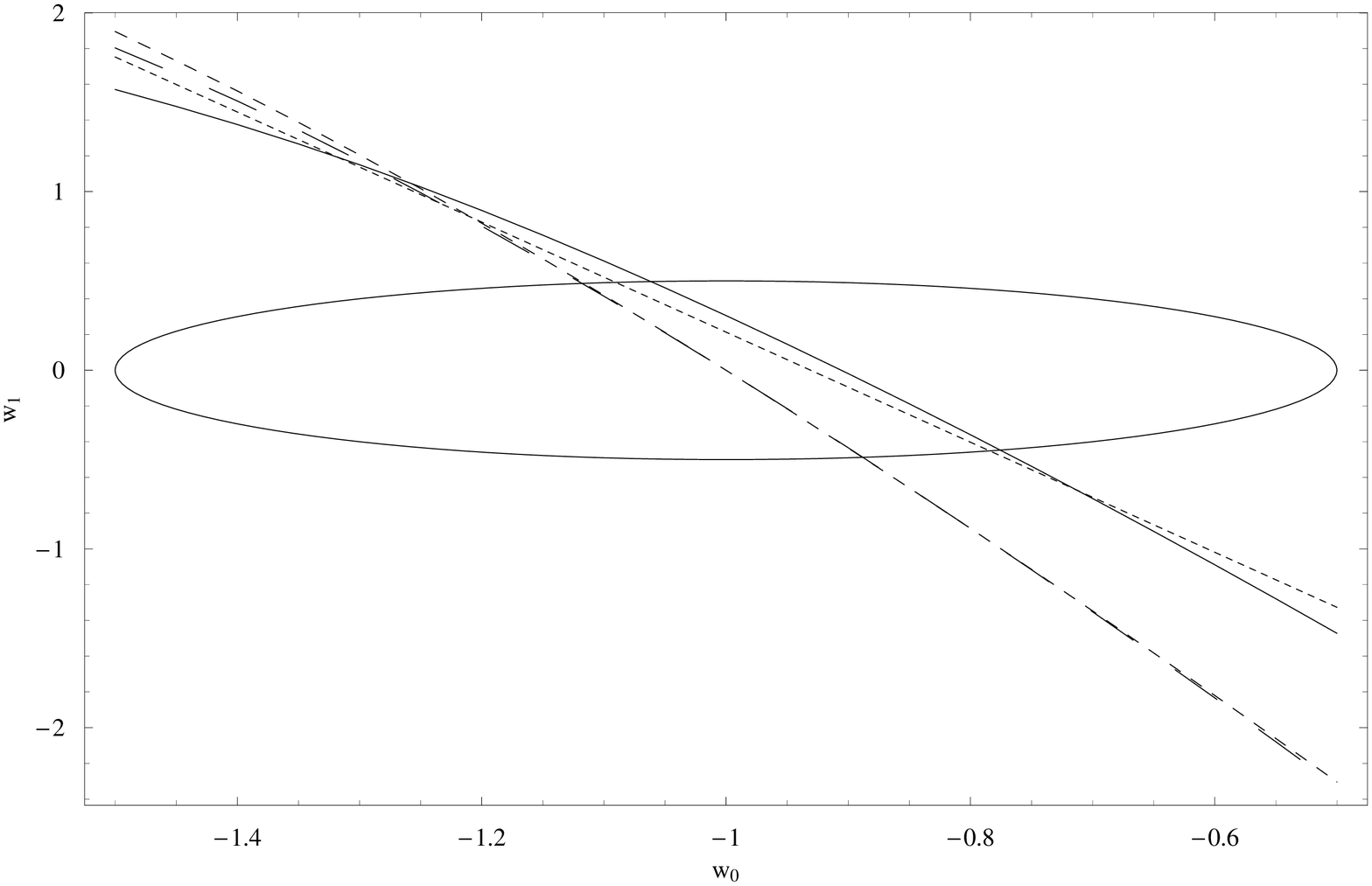}} \caption{The locus of
$w_0$, $w_1$ for Case 1 (solid line), Case 2 (dotted line), Case 3
(circle, see text), Case 4 ($m=2$, $n=0$ short dashed line) and
Case 4 ($m=1$, $n=1$ long dashed line).}
\end{figure*}

\begin{itemize}
\item Geometric methods probe the large scale geometry of
space-time directly through the redshift dependence of
cosmological distances [$d_L(z)$ or $d_A(z)$]. They thus determine
$H(z)$ independent of the validity of Einstein equations.

\item Dynamical methods determine $H(z)$ by measuring the
evolution of energy density (background or perturbations) and
using a gravity theory to relate them with geometry ie with
$H(z)$. These methods rely on knowledge of the dynamical equations
that connect geometry with energy and may therefore be used in
combination with geometric methods to test these dynamical
equations.
\end{itemize}

A very accurate and deep geometrical probe of dark energy is the
angular scale of the sound horizon at the last scattering surface as
encoded in the location $l_1^{TT}$ of the first peak of the
Cosmic Microwave Background (CMB)
temperature perturbation spectrum. This probe is described by the so
called CMB shift parameter (cf. Bond, Efstathiou \& Tegmark 1997;
Trotta 2004; Nesseris \& Perivolaropoulos 2007) which is defined as
\be R=\frac{l_1^{'TT}}{l_1^{TT}} \ee where $l_1^{TT}$ is the temperature
perturbation CMB spectrum multipole of the first acoustic peak. In
the definition of $R$, $l_1^{TT}$ corresponds to the model (with
fixed $\Omega_{\rm m}$, $\Omega_{\rm b}$ and $h$) characterized by the shift
parameter and $l_1^{'TT}$ to a \textit{reference} flat SCDM model
($\Omega_{\rm m}= 1$) with the same $\omega_{m} =\Omega_{\rm m} h^2$
and $\omega_{b} =
\Omega_{\rm b} h^2$ as the original model.
Recently, Nesseris \& Perivolaropoulos (2007) have found
that models based on
general relativity that have identical shift parameter $R$ and
matter density $\Omega_{\rm m}$ also lead to almost
identical ISW effect despite of their possible differences in
the cosmic expansion histories.

The aim of this work is to
investigate our suspicion that the (geometrical) CMB
shift parameter is somehow
associated with the (dynamical) fluctuation growth rate
in the context of general relativity. Note,
that a possible violation of this connection may
thus be viewed as a hint for modifications of general relativity.
The structure of the paper is as follows.
The basic theoretical elements are presented in section 2.
The results are presented in
section 3 by solving numerically the time evolution equation
for the mass density contrast for various
flat dark energy models that share the same value of shift
parameter and value of $\Omega_{\rm m}$. In section 4 we
draw our conclusions. Finally, in the appendix we have treated
analytically, up to a certain point,
the differential equation for the mass density contrast
considering different dark energy models with a
time varying equation of state.

\section{Theoretical elements}
The location $l_1^{TT}$
of the first acoustic peak in the CMB temperature spectrum can be connected with the angular
diameter distance $d_A$ and with the sound horizon $r_s$ both at the
last scattering surface ($z = z_{ls}$) and then the shift parameter
(see Nesseris \& Perivolaropoulos 2007 and references therein for details)
can be brought to the form $R'=\frac{2}{\sqrt{\Omega_{\rm m}}\int_0^{z_{ls}}
\frac{dz}{E(z)}}$. The expression usually used for the shift
parameter is \be R=\sqrt{\Omega_{\rm m}}\int_{a_{ls}}^1 \frac{da}{a^2
H(a)/H_0}=\sqrt{\Omega_{\rm m}}\int_0^{z_{ls}} \frac{dz}{E(z)} \label{shift}
\ee where $E(z)\equiv H(z)/H_0$.

In order to define $E(z)$ we use the Chevalier-Polarski-Linder
(CPL)[Chevallier \& Polarski 2001; Linder 2003]
parametrization for
which \be w(a)=w_0+w_1(1-a) \label{wcpl} \ee and \be E^{2}(a)=
\Omega_{\rm m} a^{-3}
+\frac{(1-\Omega_{\rm m})a^{-3}}{f(a)} \ee
where
\be
f(a)={\rm exp}\left[-3 \int_{a}^{1} \frac{w(u)}{u} du \right]=
a^{3(w_{0}+w_{1})} e^{-3w_{1}(a-1)} \;\;.
\ee

\begin{figure*}
\mbox{\hspace{-1cm} \epsfxsize=18cm \epsfysize=5cm
\epsffile{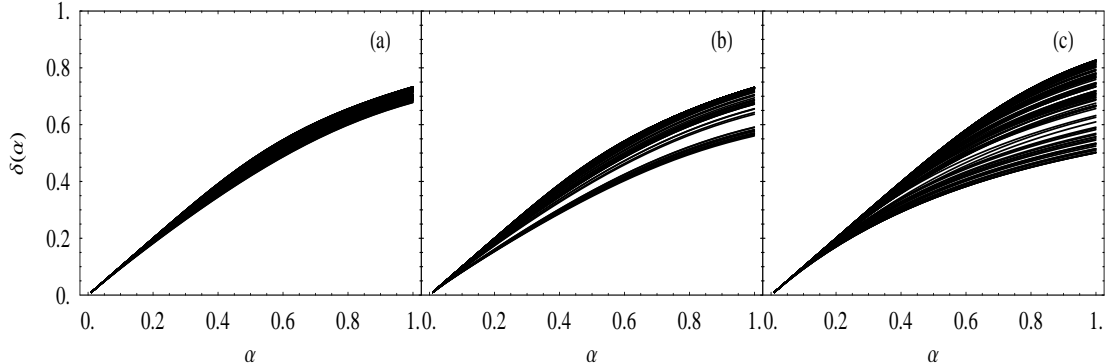}} \caption{ The growth factor for the first
three cases and all Monte-Carlo values of $w_0$ and $w_1$
($\Omega_{\rm m} =0.25$). Case 1 corresponds to Fig.2a and cases 2
and 3 to Fig.2b and Fig.2c respectively. Note that for this figure
we do not normalize the growth factor with its value at $a=1$ as
we are interested in its range of values at $z=0$.}
\end{figure*}

On the other hand, a dynamical probe of geometry is the measured linear growth factor
of the matter density perturbations $\delta(a)$. The evolution
equation of the growth factor for models where the dark energy
fluid has a vanishing anisotropic stress and the matter fluid is not
coupled to other matter species is given by
(Peebles 2003; Stabenau \& Jain 2006; Uzan 2007): \be
\ddot{\delta}+2H(t)\dot{\delta}-4\pi G \rho_{\rm m} \delta=0
\label{deltatime1} \ee where dots denote derivatives with respect to
time. Useful expressions of the growth factor can be found for the
$\Lambda$CDM cosmology in Peebles (1993) and for 
the quintessence scenario ($w=const$) in Silveira \& Waga (1994), 
Wang \& Steinhardt (1998), Basilakos (2003), 
Nesseris \& Perivolaropoulos (2008) and for the 
scalar tensor models in Gannouji \& Polarski (2008).
As an example, in the case of $\Lambda$CDM cosmology the growth factor is
of the form \be
\delta(a)=\frac{5\Omega_{\rm m}}{2}\frac{H(a)}{H_0}\int_0^a
\frac{da'}{(a' H(a')/H_0)^3} \label{grdod} \ee which for $a=1$ has
some similarity with the form of the shift parameter (see eq.\ref{shift}).
Finally, it is interesting to mention here that for
dark energy models with a time varying equation of state,
an analytical solution for $\delta(a)$ has yet to be found because
the basic differential equation (\ref{deltatime1}) becomes
more complicated than in models with constant $w$. However, in a
recent paper (Linder \& Cahn 2007) a growth index $\gamma$ was used
to parameterize the linear growing mode including models with a time
varying equation of state (see next section).

\section{The connection between the CMB shift parameter and the growth factor}
In order to explore the above mentioned connection of the CMB
shift parameter $R$ of eq. (\ref{shift}) with the growth factor
$\delta(a)$, we perform a Monte-Carlo analysis in the parameter
space of $w_0 - w_1$ of the parametrization (\ref{wcpl}). In
particular we fix $\Omega_{\rm m}$ (for example 0.25) and compare the 
variation of the growth
factor for $w_0-w_1$ pairs corresponding to fixed $R$ with the
corresponding variation when other combinations of $w_0-w_1$ are
fixed. Specifically, we consider and compare 4 cases:

\begin{itemize}
    \item Case 1: $w_0$ and $w_1$ are constrained by a fixed value
    of the CMB shift parameter, $R(w_0 , w_1)=1.7$
    \item Case 2: $w_0$ and $w_1$ are constrained by a linear
    relation of the form $w_1=-2.9-3.1 w_0$ approximating the locus of
    the $w_0$ and $w_1$ that satisfy the relation $R(w_0 ,
    w_1)=1.7$ (see Fig.1). This approximation is accurate to within
    about $5\%$ and provides an estimate of the uncertainties
    introduced in the predicted value of the growth factor if the
    shift parameter is not accurately measured.
    \item Case 3: $w_0$ and $w_1$ are constrained to be on a
    circle of radius 0.5 and center the $\Lambda$CDM point (-1,0)
    \item Case 4: $w_0$ and $w_1$ are constrained by a fixed value of an integral ansatz
    of a form similar to the CMB shift parameter \be
    A(w_0,w_1)_{m,n}=\sqrt{\Omega_{\rm m}}\int_0^{z_{ls}} \frac{dz}{(1+z)^n
    E^{m}(z)}\ee
for various values of the parameters $m,n$ (see Table 1). Note
that for $(m,n)=(1,0)$ we get the usual CMB shift parameter.
\end{itemize}

In the first two cases $w_0$ is a random variable in the range
$[-1.5,-0.5]$ and for each value of $w_0$ we use the constraint
equation to solve for $w_1$. In case 3, $w_0$ and $w_1$ are on a
circle of radius 0.5 and center the $\Lambda$CDM point (-1,0). In
Fig.1 we show the locus of the points $w_0$ and $w_1$ that satisfy
the constraints of cases 1-4. Notice that in Fig. 1, case 3 is shown
as an ellipse and not a circle due to the fact that the aspect ratio
is chosen so that the loci for the other cases are shown optimally.
Also, in Fig.2 we present the growth factor 
evolution which is derived by solving
numerically eq. (\ref{deltatime1}), for the first three cases and
all Monte-Carlo values of $w_0$ and $w_1$ (100 pairs each time). In
case 1 the values of $\delta(a=1)$ are much more constrained than
the other two cases as the range (dispersion) of the growth factor
at $a=1$, ie $ Max[\delta(a=1)]-Min[\delta(a=1)]$ for the first case
is $0.055$ (corresponding to a variation of the mean value of
$\delta(a=1)$ of less than $8\%$) while in the second and third
cases it is $3.1$ \textit{($33\%$ variation)} and $5.9$
\textit{($49\%$ variation)} times that value. We have also
investigated the sensitivity of our analysis to the matter density
parameter. In particular, we confirmed that in the range
$\Omega_{\rm m} \in [0.2,0.3]$ our results depend weakly on the
value of $\Omega_{\rm m}$. In fact, the present time dispersion of
the growth factor for fixed shift parameter varies from $3.5 \%$ for
$\Omega_{\rm m} =0.2$ to $9\%$ for $\Omega_{\rm m} =0.3$  ($8\%$ for
$\Omega_{\rm m} =0.25$). Thus our main result persists for all
physical values of $\Omega_{\rm m}$ and it strongly indicates that the CMB
shift parameter is somehow {\it associated} with the growth factor
in the context of general relativity.

The linear relation of the form $w_1=-2.9-3.1 w_0$
corresponding to Fig. 2b provides a rough approximation (good to
about $5\%$) of the locus of points that satisfy the relation
$R(w_0,w_1)=1.7$. This introduces significant additional dispersion
to the present day growth factor (the dispersion goes to $33\%$ in
Fig. 2b from the $8\%$ obtained with the exact locus of fixed $R$ in
Fig. 2a). Once we improve the $w_1$ expression with an appropriate
quadratic term in $w_0$, the approximation improves from about $5\%$
to about $0.3\%$ and the growth factor dispersion drops back to
$8.5\%$ (almost the same as with the exact locus of fixed $R$). This
is an interesting result as it means that the shift parameter should
be measured with $1\sigma$ errors better than $1\%$ for a
determination of the growth factor to an accuracy better than
$10\%$. This accuracy of measurement of the shift parameter however
is not far from present day measurements which have determined $R$
to within $1.5\%$ ($R=1.7\pm 0.03$ from Wang \& Mukherjee 2007).

\begin{figure*}
\mbox{\epsfxsize=16cm \epsfysize=6cm \epsffile{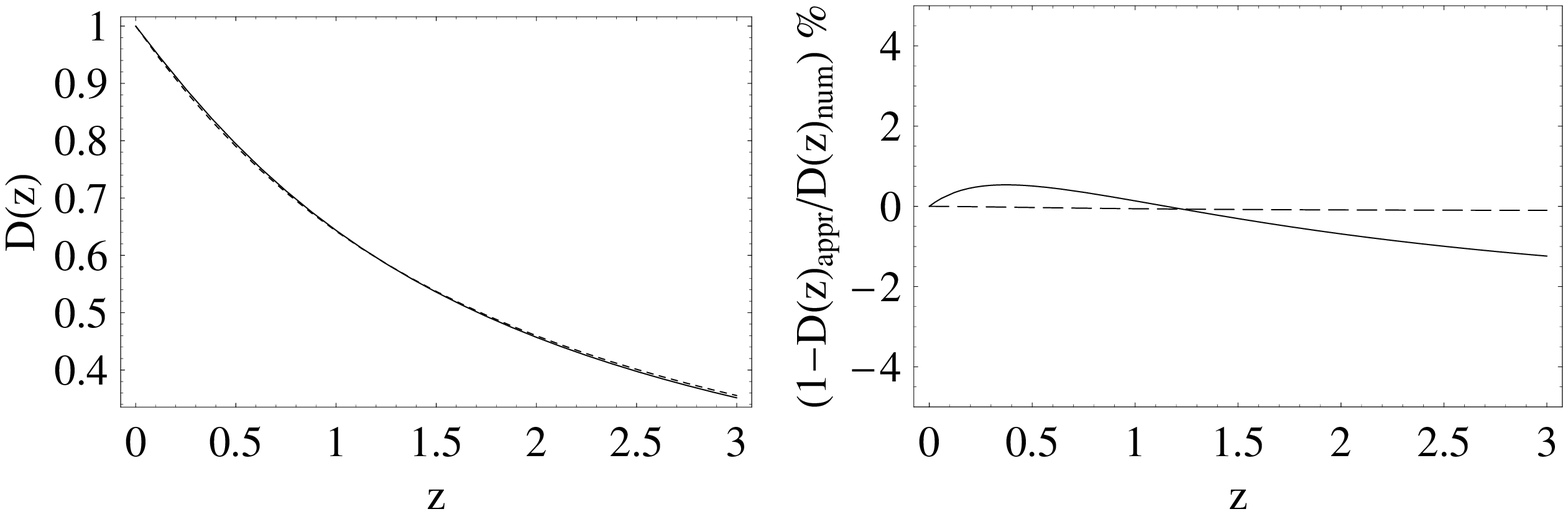}}
\caption{{\it Left panel}: The growth factor as a function of
redshift. The solid line represents our analytical approximation
(see (\ref{gmodesol})), while the dashed line represents the
parametrized growth factor (see (\ref{glind}) derived by Linder \&
Cahn (2007)). {\it Right panel}: The percent accuracy
$(1-D_{appr}/D_{num})\%$ of the two approximations, eq.
(\ref{glind}) (dashed line) and eq. (\ref{gmodesol}) (solid line).
Note, that we use $\Omega_{\rm m}=0.25$ and
$(w_0,w_1)=(-0.95,0.43)$.}
\end{figure*}

In case 4 the analysis is similar to case 1, ie $w_0$ is a random
variable but now $w_1$ is found from the generalized constraint \be
A(w_0,w_1)_{m,n}=A_{fixed}~_{m,n} \label{constr}\ee where
$A_{fixed}~_{m,n}$ is the value of $A(w_0,w_1)_{m,n}$ for a
$\Lambda$CDM cosmology $(w_0=-1,w_1=0)$ for the respective values of
$m,n$. We computed the range (dispersion) of the growth factor at
$a=1$, ie $ Max[\delta(a=1)]-Min[\delta(a=1)]$ for the values $w_0$
and $w_1$ derived from the constraint eq. (\ref{constr}). In Table 1
we show the ratio of these values for various $m,n$ to the value of
the range of the growth factor for $m=1,n=0$ [when
$A(w_0,w_1)_{m,n}$ goes over to the CMB shift parameter $R$],ie \be
\frac{ (Max[\delta(a=1)]-Min[\delta(a=1)])_{(m,n)}}{
(Max[\delta(a=1)]-Min[\delta(a=1)])_{(1,0)}} \label{4table1}\;\;.\ee 
Finally, we checked that random values for both $w_0$ and $w_1$ give
a much larger dispersion on the values of $\delta(a=1)$, as exactly
they should. Notice that the CMB shift parameter $R$ ($m=1,n=0$)
seems to constrain the growth factor by about an order of magnitude
or more compared to other forms of the integral ansatz (different
values of $m$ and $n$). From a theoretical point of view a
possible relation between the CMB shift parameter and the growth
factor can be used as a viable test for the general relativity. As
expected (Bertschinger 2006), changing the validity of Einstein's
field equations (the so called theory of modified gravity), we
change accordingly the growth factor [see Gannouji \& Polarski
(2008) for a detailed investigation of the growth factor in
scalar-tensor theories]. In contrast the behavior of the CMB shift
parameter remains unaltered, simply because the latter is a
geometrical function (see e.g. Nesseris \& Perivolaropoulos 2007).
Thus, a mismatch between the measured value of the shift parameter
and the measured value of the linear growth factor would be a hint
towards modified gravity.

\begin{table}
{\small
\caption[]
{The ratio of the values of the range of
the growth factor at $a=1$ for various $m,n$ to that for
$m=1,n=0$. The case $m=1$, $n=0$ corresponds to the CMB shift
parameter. Notice that for these values of $m,n$ the ratio becomes
minimal.}
\tabcolsep 6pt
\begin{tabular}{cccccc}
\hline
$m/n$& 0 & 1& 2& 3& 4 \\ \hline
0&  9.64& 9.74& 9.75&  9.16&9.79\\
1&  1.00& 5.92& 10.71&  11.72&54.24\\
2&  7.95& 10.84& 11.58&  56.81& 376.47\\
3&  10.89& 11.89& 69.62&  428.33&2969.39\\
4& 11.61& 73.53& 247.44&  1422.99& 20633.90\\ \hline
\end{tabular}
}
\end{table}

Verifying this connection between the growth factor and the CMB
shift parameter $R$ analytically requires an exact or approximate
solution to the differential equation for the evolution of density
perturbations when evolving dark energy is taken into account. In
what follows we discuss some recent attempts towards the
construction of such approximate solutions.

A well known approximate solution to eq.(\ref{deltatime1}) is found
by Linder \& Cahn (2007), where a growth index $\gamma$ was used
to parameterize the linear growing mode for models with a time
varying equation of state. Specifically, the growth index $\gamma$
was defined through \be
D(a)={\rm exp} \left[\int_{1}^{a} \frac{\Omega_{\rm m}^{\gamma}(u)}{u} {\rm d}u \right]
\label{glind} \ee
where $D(a)$ is the growth factor normalized to unity and it was
found that $\gamma$ can be approximated by \be
\gamma=\frac{6-3(1+w_{\infty})}{11-6(1+w_{\infty})} \ee where
$w_{\infty} \equiv w(z >> 1)$.

The previous approach was based on the approximation that the
universe is not too far from being matter-dominated. However, by
utilizing just some basic elements from the differential equation
theory we have solved analytically up to a certain point
eq.(\ref{deltatime1}) (see the Appendix for details), assuming
that the equation of state parameter is a function of time.
In this approach the growing
$D_{+}(a)$ and decaying $D_{-}(a)$ modes are given by \be
D_{\pm}(a)\simeq a^{-3/2}E^{-1/2}(a) \;{\rm
exp}\left(\mp\frac{\sqrt{21}}{3} \int_{1}^{a} |g(u)|^{1/2}{\rm d}u
\right) \label{gmodesol} \ee
where $ g(a)$ is defined in the Appendix.

Utilizing the best-fit cosmological parameters $\Omega_{\rm
m}=0.25$ and $(w_0,w_1)=(-0.95,0.43)$ obtained from the Gold06
SNIa dataset and the CMB shift parameter, in the right panel of
Fig.3, we present the approximated growth factor (left panel), as
a function of redshift by utilizing eq. (\ref{gmodesol}) (solid
line) and eq. (\ref{glind}) respectively (dashed line). It is
obvious that our analytical approximation is indeed close to that
found by Linder \& Cahn (2007). In the right panel of Fig. 3 we
show the percent accuracy $(1-D_{appr}/D_{num})\%$ of the two
approximations $D{appr}$, compared to the numerical solution. Eq.
(\ref{glind}) deviates from the numerical solution by $0.17\%$
while our $D_{+}(z)$ approximation deviates by $0.5-1.1\%$.

\section{Conclusions}
In this work, we found that flat models which have
the same $\Omega_{\rm m}$ and
CMB shift parameter $R$ but different $H(a)$ and $w(a)$ also have
very similar growth of perturbations even though the dark energy
density evolution is quite different. This was done by comparing
various forms of constraints for $w_0$ and $w_1$, besides the CMB
shift parameter, by a Monte-Carlo simulation. In all cases
considered, models constrained by the CMB shift parameter had also a
very similar growth factor.

\section*{Appendix}
In this appendix we try to treat analytically, as much as possible,
the problem of the growth factor evolution for dark energy models
with a time varying equation of state.

The time evolution equation for the mass density contrast, modeled
as a pressureless fluid, is obtained from the Euler and matter
stress energy conservation equations as: \be
\ddot{\delta}+2H(t)\dot{\delta}-4\pi G \rho_{\rm m} \delta=0\;\; ,
\label{deltatime} \ee where dots denote derivatives with respect to
time. This differential equation is valid for models where the dark
energy fluid has a vanishing anisotropic stress and the matter fluid
is not coupled to other matter species, however
see Uzan (2007) for a detailed discussion of the modifications
that appear on the right-hand side of the above equation
when such
terms are present. Changing variables from $t$ to $a$ the above
equation becomes (see also Linder 2003) \be
\delta^{''}+A(a)\delta^{'}+B(a)\delta=0 \ \label{odedelta} \ee where
$ A(a)= \frac{3}{2a}\left[1-\frac{(1-\Omega_{\rm m})w(a)}
{[1-\Omega_{\rm m}+\Omega_{\rm m} f(a)]}
\right]$ and $B(a)=-\frac{3}{2a^{2}}
\left[\frac{\Omega_{\rm m} f(a)}{1-\Omega_{\rm m}+
\Omega_{\rm m} f(a)}\right]\;\;$.

Performing now the following transformation \be \delta(a)=y(a) \;{\rm
exp} \left[-\frac{1}{2} \int_{1}^{a} A(u){\rm d}u\right] \ee the
linear mass density fluctuations is written: \be \delta(a)=y(a)\;
a^{-3/2}E^{-1/2}(a) \;\; \label{sol11}\ee
and the unknown function $y$ satisfies
the following differential equation: \be y^{''}-g(a)y=0
\label{yode}\ee with a relevant factor of \be
g(a)=\frac{1}{2}A^{'}(a)+\frac{1}{4}A^{2}(a)-B(a) \;\;. \ee
It becomes evident, that a major part of the
pure solution is described by the expression
$a^{-3/2}E^{-1/2}(a)$.

Of course, in order to solve fully the problem we have to derive the functional
form of $y$. In particular, we write eq. (\ref{yode}) as follows \be
\frac{y^{'2}}{2}-\frac{g(a)y^{2}}{2}+\int_{1}^{a} \frac{y^{2}(u)}{2}
g^{'}(u) {\rm d}u=-c_{1} \ee or \be |g(a)y^{2}|=|c^{2} +\int_{1}^{a}
y^{2}(u) g^{'}(u) {\rm d}u+y^{'2}| \;\;\;\;\;\;(c^{2}=2c_{1})\;\;.
\ee
From a mathematical point of view we can select the integration
constant $c$ to be large enough such as \be n^{2}|g(a)|y^{2}\le
n^{2}c^{2}+ \int_{1}^{a} n^{2}|g(u)|y^{2}(u)
\frac{|g^{'}(u)|}{|g(u)|} {\rm d}u+y^{'2} \label{ineq1} \ee where
$n^{2}$ is the normalization constant of the problem. Now, we can
use Gronwall's theorem (see Gronwall 1919), which is a well known
theorem from the differential equation theory.

{\it Theorem}: Lets assume that $\mu : [a,\beta] \rightarrow
[0,\infty)$ and $y : [a,\beta] \rightarrow [0,\infty)$ continuous
functions and $\lambda \in {\cal R}$. If \be y \le \lambda +
|\int_{t_{0}}^{t} \mu(x) y(x) {\rm d}x | \;\;\;\; \forall t\in
[a,\beta] \ee then \be y \le \lambda \;{\rm exp}
\left(|\int_{t_{0}}^{t} \mu(x) {\rm d}x | \right)\;\;\;\; \forall
t\in [a,\beta] \ee

\noindent and using it on (\ref{ineq1}) we get the following useful
formula: \be n^{2}|g(a)|y^{2} \le n^{2}c^{2} {\rm exp} \left(
\int_{1}^{a} \frac{|g^{'}(u)|}{|g(u)|} {\rm d}u \right)+y^{'2} \;\;.
\ee Doing so it turns out that a possible approximation could be:
\be n^{2}|g(a)|y^{2} \simeq n^{2}c^{2}|g(a)|+y^{'2} \ee from which
we get that: \be y(a)\simeq c \;\;{\rm cosh}\left(\pm n\int_{1}^{a}
|g(u)|^{1/2}{\rm d}u \right) \;\;. \label{yresult}\ee Taking into
account eq. (\ref{sol11}), we can obtain the following approximation: \be
\delta(a)\simeq c\; a^{-3/2}E^{-1/2}(a)\; {\rm cosh}\left(\pm
n\int_{1}^{a} |g(u)|^{1/2}{\rm d}u \right)\;\;. \ee

In order to normalize our analytical expression we use as a limiting
case the Einstein de-Sitter model 
($\Omega_{\rm m}=1$, $g(a)=21/16a^{2}$) in which the behavior
of the corresponding growing mode is well known $D_{+}(a)=a$.
Indeed, doing so we get  $n=\sqrt{21}/3$ and thus, the following
normalized growing $D_{+}$ and decaying $D_{-}$ modes respectively
 become:\be D_{\pm}(a)\simeq a^{-3/2}E^{-1/2}(a) \;{\rm
exp}\left(\mp\frac{\sqrt{21}}{3} \int_{1}^{a} |g(u)|^{1/2}{\rm d}u
\right) \label{gmodesol1} \ee

\end{document}